\documentclass[12pt]{article}
\usepackage{xspace}
\usepackage{amssymb}

\def\babs{\begin{abstract}}             \def\eabs{\end{abstract}}
\def\barr{\begin{array}}                \def\earr{\end{array}}
\def\bcc{\begin{center}}                \def\ecc{\end{center}}

\def\bdes{\begin{description}}          \def\edes{\end{description}}
\def\bdoc{\begin{document}}             \def\edoc{\end{document}}
\def\ben{\begin{enumerate}}             \def\een{\end{enumerate}}
\def\beqn{\begin{eqnarray}}             \def\eeqn{\end{eqnarray}}
\def\beqnl#1{\beqn\label{#1}}           \def\eeqnl#1{\label{#1}\eeqn}
\def\beqnx{\begin{eqnarray*}}           \def\eeqnx{\end{eqnarray*}}
\def\bseqn{\begin{subeqnarray}}         \def\eseqn{\end{subeqnarray}}
\def\beq#1\eeq{\begin{equation}#1\end{equation}}
\def\bal#1\eal{\begin{align}#1\end{align}}
\def\balx#1\ealx{\begin{align*}#1\end{align*}}
\def\beqx{$$}                           \def\eeqx{$$}
\def\bfig{\protect\begin{figure}}       \def\efig{\protect\end{figure}}
\def\bfigx{\protect\begin{figure*}}     \def\efigx{\protect\end{figure*}}
\def\bfl{\begin{flushleft}}             \def\efl{\end{flushleft}}
\def\bfr{\begin{flushright}}            \def\efr{\end{flushright}}
\def\bit{\begin{itemize}}               \def\eit{\end{itemize}}
\def\bpic{\begin{picture}}              \def\epic{\end{picture}}
\def\bqu{\begin{quote}}                 \def\equ{\end{quote}}
\def\bqun{\begin{quotation}}            \def\equn{\end{quotation}}
\def\bsl{\begin{slide}}                 \def\esl{\end{slide}}
\def\btabb{\begin{tabbing}}             \def\etabb{\end{tabbing}}
\def\btabl{\begin{table}}               \def\etabl{\end{table}}
\def\btablx{\begin{table*}}             \def\etablx{\end{table*}}
\def\btab{\begin{tabular}}              \def\etab{\end{tabular}}
\def\btabu{\begin{tabular}}             \def\etabu{\end{tabular}}
\def\btabx{\begin{tabular*}}            \def\etabx{\end{tabular*}}
\def\bbib{\begin{thebibliography}{999}} \def\ebib{\end{thebibliography}}
\def\bver{\begin{verbatim}}             \def\ever{\end{verbatim}}

\def\XS{\xspace}

\def\bm#1{\mbox{\boldmath $#1$}}                

\def\sbm#1{\mbox{\boldmath $#1$}}
\def\sbmp#1{\mbox{\boldmath{$\scriptstyle #1$}}}
\def\sdmp#1{\mbox{\scriptsize #1}}
\def\trans#1{\mbox{$#1^{\sdmp t}$}}
\def\sbv#1{\mbox{\bf #1}}
\def\sbvp#1{\mbox{\scriptsize {\bf #1}}}
\def\sdm#1{\mbox{#1}}

\def\Ab{{\sbm A}\XS}    \def\ab{{\sbm a}\XS}
\def\Bb{{\sbm B}\XS}    \def\bb{{\sbm b}\XS}
\def\Cb{{\sbm C}\XS}    \def\cb{{\sbm c}\XS}
\def\Db{{\sbm D}\XS}    \def\db{{\sbm d}\XS}
\def\Eb{{\sbm E}\XS}    \def\eb{{\sbm e}\XS}
\def\Fb{{\sbm F}\XS}    \def\fb{{\sbm f}\XS}
\def\Gb{{\sbm G}\XS}    \def\gb{{\sbm g}\XS}
\def\Hb{{\sbm H}\XS}    \def\hb{{\sbm h}\XS}
\def\Ib{{\sbm I}\XS}    \def\ib{{\sbm i}\XS}
\def\Jb{{\sbm J}\XS}    \def\jb{{\sbm j}\XS}
\def\Kb{{\sbm K}\XS}    \def\kb{{\sbm k}\XS}
\def\Lb{{\sbm L}\XS}    \def\lb{{\sbm l}\XS}
\def\Mb{{\sbm M}\XS}    \def\mb{{\sbm m}\XS}
\def\Nb{{\sbm N}\XS}    \def\nb{{\sbm n}\XS}
\def\Ob{{\sbm O}\XS}    \def\ob{{\sbm o}\XS}
\def\Pb{{\sbm P}\XS}    \def\pb{{\sbm p}\XS}
\def\Qb{{\sbm Q}\XS}    \def\qb{{\sbm q}\XS}
\def\Rb{{\sbm R}\XS}    \def\rb{{\sbm r}\XS}
\def\Sb{{\sbm S}\XS}    \def\sb{{\sbm s}\XS}
\def\Tb{{\sbm T}\XS}    \def\tb{{\sbm t}\XS}
\def\Ub{{\sbm U}\XS}    \def\ub{{\sbm u}\XS}
\def\Vb{{\sbm V}\XS}    \def\vb{{\sbm v}\XS}
\def\Wb{{\sbm W}\XS}    \def\wb{{\sbm w}\XS}
\def\Xb{{\sbm X}\XS}    \def\xb{{\sbm x}\XS}
\def\Yb{{\sbm Y}\XS}    \def\yb{{\sbm y}\XS}
\def\Zb{{\sbm Z}\XS}    \def\zb{{\sbm z}\XS}

\def\Abt{\trans{\sbm A}\XS}	\def\abt{\trans{\sbm a}\XS}
\def\Bbt{\trans{\sbm B}\XS}	\def\bbt{\trans{\sbm b}\XS}
\def\Cbt{\trans{\sbm C}\XS}	\def\cbt{\trans{\sbm c}\XS}
\def\Dbt{\trans{\sbm D}\XS}	\def\dbt{\trans{\sbm d}\XS}
\def\Ebt{\trans{\sbm E}\XS}	\def\ebt{\trans{\sbm e}\XS}
\def\Fbt{\trans{\sbm F}\XS}	\def\fbt{\trans{\sbm f}\XS}
\def\Gbt{\trans{\sbm G}\XS}	\def\gbt{\trans{\sbm g}\XS}
\def\Hbt{\trans{\sbm H}\XS}	\def\hbt{\trans{\sbm h}\XS}
\def\Ibt{\trans{\sbm I}\XS}	\def\ibt{\trans{\sbm i}\XS}
\def\Jbt{\trans{\sbm J}\XS}	\def\jbt{\trans{\sbm j}\XS}
\def\Kbt{\trans{\sbm K}\XS}	\def\kbt{\trans{\sbm k}\XS}
\def\Lbt{\trans{\sbm L}\XS}	\def\lbt{\trans{\sbm l}\XS}
\def\Mbt{\trans{\sbm M}\XS}	\def\mbt{\trans{\sbm m}\XS}
\def\Nbt{\trans{\sbm N}\XS}	\def\nbt{\trans{\sbm n}\XS}
\def\Obt{\trans{\sbm O}\XS}	\def\obt{\trans{\sbm o}\XS}
\def\Pbt{\trans{\sbm P}\XS}	\def\pbt{\trans{\sbm p}\XS}
\def\Qbt{\trans{\sbm Q}\XS}	\def\qbt{\trans{\sbm q}\XS}
\def\Rbt{\trans{\sbm R}\XS}	\def\rbt{\trans{\sbm r}\XS}
\def\Sbt{\trans{\sbm S}\XS}	\def\sbt{\trans{\sbm s}\XS}
\def\Tbt{\trans{\sbm T}\XS}	\def\tbt{\trans{\sbm t}\XS}
\def\Ubt{\trans{\sbm U}\XS}	\def\ubt{\trans{\sbm u}\XS}
\def\Vbt{\trans{\sbm V}\XS}	\def\vbt{\trans{\sbm v}\XS}
\def\Wbt{\trans{\sbm W}\XS}	\def\wbt{\trans{\sbm w}\XS}
\def\Xbt{\trans{\sbm X}\XS}	\def\xbt{\trans{\sbm x}\XS}
\def\Ybt{\trans{\sbm Y}\XS}	\def\ybt{\trans{\sbm y}\XS}
\def\Zbt{\trans{\sbm Z}\XS}	\def\zbt{\trans{\sbm z}\XS}

\def\Abp{{\sbmp A}\XS}		\def\abp{{\sbmp a}\XS}
\def\Bbp{{\sbmp B}\XS}		\def\bbp{{\sbmp b}\XS}
\def\Cbp{{\sbmp C}\XS}		\def\cbp{{\sbmp c}\XS}
\def\Dbp{{\sbmp D}\XS}		\def\dbp{{\sbmp d}\XS}
\def\Ebp{{\sbmp E}\XS}		\def\ebp{{\sbmp e}\XS}
\def\Fbp{{\sbmp F}\XS}		\def\fbp{{\sbmp f}\XS}
\def\Gbp{{\sbmp G}\XS}		\def\gbp{{\sbmp g}\XS}
\def\Hbp{{\sbmp H}\XS}		\def\hbp{{\sbmp h}\XS}
\def\Ibp{{\sbmp I}\XS}		\def\ibp{{\sbmp i}\XS}
\def\Jbp{{\sbmp J}\XS}		\def\jbp{{\sbmp j}\XS}
\def\Kbp{{\sbmp K}\XS}		\def\kbp{{\sbmp k}\XS}
\def\Lbp{{\sbmp L}\XS}		\def\lbp{{\sbmp l}\XS}
\def\Mbp{{\sbmp M}\XS}		\def\mbp{{\sbmp m}\XS}
\def\Nbp{{\sbmp N}\XS}		\def\nbp{{\sbmp n}\XS}
\def\Obp{{\sbmp O}\XS}		\def\obp{{\sbmp o}\XS}
\def\Pbp{{\sbmp P}\XS}		\def\pbp{{\sbmp p}\XS}
\def\Qbp{{\sbmp Q}\XS}		\def\qbp{{\sbmp q}\XS}
\def\Rbp{{\sbmp R}\XS}		\def\rbp{{\sbmp r}\XS}
\def\Sbp{{\sbmp S}\XS}		\def\sbp{{\sbmp s}\XS}
\def\Tbp{{\sbmp T}\XS}		\def\tbp{{\sbmp t}\XS}
\def\Ubp{{\sbmp U}\XS}		\def\ubp{{\sbmp u}\XS}
\def\Vbp{{\sbmp V}\XS}		\def\vbp{{\sbmp v}\XS}
\def\Wbp{{\sbmp W}\XS}		\def\wbp{{\sbmp w}\XS}
\def\Xbp{{\sbmp X}\XS}		\def\xbp{{\sbmp x}\XS}
\def\Ybp{{\sbmp Y}\XS}		\def\ybp{{\sbmp y}\XS}
\def\Zbp{{\sbmp Z}\XS}		\def\zbp{{\sbmp z}\XS}

\def\Ac{\mbox{$\cal A$}\XS}	
\def\Bc{\mbox{$\cal B$}\XS}	
\def\Cc{\mbox{$\cal C$}\XS}	
\def\Dc{\mbox{$\cal D$}\XS}	
\def\Ec{\mbox{$\cal E$}\XS}	
\def\Fc{\mbox{$\cal F$}\XS}	
\def\Gc{\mbox{$\cal G$}\XS}	
\def\Hc{\mbox{$\cal H$}\XS}	
\def\Ic{\mbox{$\cal I$}\XS}	
\def\Jc{\mbox{$\cal J$}\XS}	
\def\Kc{\mbox{$\cal K$}\XS}	
\def\Lc{\mbox{$\cal L$}\XS}	
\def\Mc{\mbox{$\cal M$}\XS}	
\def\Nc{\mbox{$\cal N$}\XS}	
\def\Oc{\mbox{$\cal O$}\XS}	
\def\Pc{\mbox{$\cal P$}\XS}	
\def\Qc{\mbox{$\cal Q$}\XS}	
\def\Rc{\mbox{$\cal R$}\XS}	
\def\Sc{\mbox{$\cal S$}\XS}	
\def\Tc{\mbox{$\cal T$}\XS}	
\def\Uc{\mbox{$\cal U$}\XS}	
\def\Vc{\mbox{$\cal V$}\XS}	
\def\Wc{\mbox{$\cal W$}\XS}	
\def\Xc{\mbox{$\cal X$}\XS}	
\def\Yc{\mbox{$\cal Y$}\XS}	
\def\Zc{\mbox{$\cal Z$}\XS}	

\def\Acp{\mbox{$\scriptstyle \cal A$}\XS}
\def\Bcp{\mbox{$\scriptstyle \cal B$}\XS}
\def\Ccp{\mbox{$\scriptstyle \cal C$}\XS}
\def\Dcp{\mbox{$\scriptstyle \cal D$}\XS}
\def\Ecp{\mbox{$\scriptstyle \cal E$}\XS}
\def\Fcp{\mbox{$\scriptstyle \cal F$}\XS}
\def\Gcp{\mbox{$\scriptstyle \cal G$}\XS}
\def\Hcp{\mbox{$\scriptstyle \cal H$}\XS}
\def\Icp{\mbox{$\scriptstyle \cal I$}\XS}
\def\Jcp{\mbox{$\scriptstyle \cal J$}\XS}
\def\Kcp{\mbox{$\scriptstyle \cal K$}\XS}
\def\Lcp{\mbox{$\scriptstyle \cal L$}\XS}
\def\Mcp{\mbox{$\scriptstyle \cal M$}\XS}
\def\Ncp{\mbox{$\scriptstyle \cal N$}\XS}
\def\Ocp{\mbox{$\scriptstyle \cal O$}\XS}
\def\Pcp{\mbox{$\scriptstyle \cal P$}\XS}
\def\Qcp{\mbox{$\scriptstyle \cal Q$}\XS}
\def\Rcp{\mbox{$\scriptstyle \cal R$}\XS}
\def\Scp{\mbox{$\scriptstyle \cal S$}\XS}
\def\Tcp{\mbox{$\scriptstyle \cal T$}\XS}
\def\Ucp{\mbox{$\scriptstyle \cal U$}\XS}
\def\Vcp{\mbox{$\scriptstyle \cal V$}\XS}
\def\Wcp{\mbox{$\scriptstyle \cal W$}\XS}
\def\Xcp{\mbox{$\scriptstyle \cal X$}\XS}
\def\Ycp{\mbox{$\scriptstyle \cal Y$}\XS}
\def\Zcp{\mbox{$\scriptstyle \cal Z$}\XS}

\def\Act{\trans{\cal A}\XS}
\def\Bct{\trans{\cal B}\XS}
\def\Cct{\trans{\cal C}\XS}
\def\Dct{\trans{\cal D}\XS}
\def\Ect{\trans{\cal E}\XS}
\def\Fct{\trans{\cal F}\XS}
\def\Gct{\trans{\cal G}\XS}
\def\Hct{\trans{\cal H}\XS}
\def\Ict{\trans{\cal I}\XS}
\def\Jct{\trans{\cal J}\XS}
\def\Kct{\trans{\cal K}\XS}
\def\Lct{\trans{\cal L}\XS}
\def\Mct{\trans{\cal M}\XS}
\def\Nct{\trans{\cal N}\XS}
\def\Oct{\trans{\cal O}\XS}
\def\Pct{\trans{\cal P}\XS}
\def\Qct{\trans{\cal Q}\XS}
\def\Rct{\trans{\cal R}\XS}
\def\Sct{\trans{\cal S}\XS}
\def\Tct{\trans{\cal T}\XS}
\def\Uct{\trans{\cal U}\XS}
\def\Vct{\trans{\cal V}\XS}
\def\Wct{\trans{\cal W}\XS}
\def\Xct{\trans{\cal X}\XS}
\def\Yct{\trans{\cal Y}\XS}
\def\Zct{\trans{\cal Z}\XS}

\def\Acb{\sbm{\cal A}\XS}
\def\Bcb{\sbm{\cal B}\XS}
\def\Ccb{\sbm{\cal C}\XS}
\def\Dcb{\sbm{\cal D}\XS}
\def\Ecb{\sbm{\cal E}\XS}
\def\Fcb{\sbm{\cal F}\XS}
\def\Gcb{\sbm{\cal G}\XS}
\def\Hcb{\sbm{\cal H}\XS}
\def\Icb{\sbm{\cal I}\XS}
\def\Jcb{\sbm{\cal J}\XS}
\def\Kcb{\sbm{\cal K}\XS}
\def\Lcb{\sbm{\cal L}\XS}
\def\Mcb{\sbm{\cal M}\XS}
\def\Ncb{\sbm{\cal N}\XS}
\def\Ocb{\sbm{\cal O}\XS}
\def\Pcb{\sbm{\cal P}\XS}
\def\Qcb{\sbm{\cal Q}\XS}
\def\Rcb{\sbm{\cal R}\XS}
\def\Scb{\sbm{\cal S}\XS}
\def\Tcb{\sbm{\cal T}\XS}
\def\Ucb{\sbm{\cal U}\XS}
\def\Vcb{\sbm{\cal V}\XS}
\def\Wcb{\sbm{\cal W}\XS}
\def\Xcb{\sbm{\cal X}\XS}
\def\Ycb{\sbm{\cal Y}\XS}
\def\Zcb{\sbm{\cal Z}\XS}

\def\Acbp{\sbmp{\cal A}\XS}
\def\Bcbp{\sbmp{\cal B}\XS}
\def\Ccbp{\sbmp{\cal C}\XS}
\def\Dcbp{\sbmp{\cal D}\XS}
\def\Ecbp{\sbmp{\cal E}\XS}
\def\Fcbp{\sbmp{\cal F}\XS}
\def\Gcbp{\sbmp{\cal G}\XS}
\def\Hcbp{\sbmp{\cal H}\XS}
\def\Icbp{\sbmp{\cal I}\XS}
\def\Jcbp{\sbmp{\cal J}\XS}
\def\Kcbp{\sbmp{\cal K}\XS}
\def\Lcbp{\sbmp{\cal L}\XS}
\def\Mcbp{\sbmp{\cal M}\XS}
\def\Ncbp{\sbmp{\cal N}\XS}
\def\Ocbp{\sbmp{\cal O}\XS}
\def\Pcbp{\sbmp{\cal P}\XS}
\def\Qcbp{\sbmp{\cal Q}\XS}
\def\Rcbp{\sbmp{\cal R}\XS}
\def\Scbp{\sbmp{\cal S}\XS}
\def\Tcbp{\sbmp{\cal T}\XS}
\def\Ucbp{\sbmp{\cal U}\XS}
\def\Vcbp{\sbmp{\cal V}\XS}
\def\Wcbp{\sbmp{\cal W}\XS}
\def\Xcbp{\sbmp{\cal X}\XS}
\def\Ycbp{\sbmp{\cal Y}\XS}
\def\Zcbp{\sbmp{\cal Z}\XS}

\def\Acbt{\trans{\sbm{\cal A}}\XS}
\def\Bcbt{\trans{\sbm{\cal B}}\XS}
\def\Ccbt{\trans{\sbm{\cal C}}\XS}
\def\Dcbt{\trans{\sbm{\cal D}}\XS}
\def\Ecbt{\trans{\sbm{\cal E}}\XS}
\def\Fcbt{\trans{\sbm{\cal F}}\XS}
\def\Gcbt{\trans{\sbm{\cal G}}\XS}
\def\Hcbt{\trans{\sbm{\cal H}}\XS}
\def\Icbt{\trans{\sbm{\cal I}}\XS}
\def\Jcbt{\trans{\sbm{\cal J}}\XS}
\def\Kcbt{\trans{\sbm{\cal K}}\XS}
\def\Lcbt{\trans{\sbm{\cal L}}\XS}
\def\Mcbt{\trans{\sbm{\cal M}}\XS}
\def\Ncbt{\trans{\sbm{\cal N}}\XS}
\def\Ocbt{\trans{\sbm{\cal O}}\XS}
\def\Pcbt{\trans{\sbm{\cal P}}\XS}
\def\Qcbt{\trans{\sbm{\cal Q}}\XS}
\def\Rcbt{\trans{\sbm{\cal R}}\XS}
\def\Scbt{\trans{\sbm{\cal S}}\XS}
\def\Tcbt{\trans{\sbm{\cal T}}\XS}
\def\Ucbt{\trans{\sbm{\cal U}}\XS}
\def\Vcbt{\trans{\sbm{\cal V}}\XS}
\def\Wcbt{\trans{\sbm{\cal W}}\XS}
\def\Xcbt{\trans{\sbm{\cal X}}\XS}
\def\Ycbt{\trans{\sbm{\cal Y}}\XS}
\def\Zcbt{\trans{\sbm{\cal Z}}\XS}

\def\AD{{\sdm A}\XS}
\def\BD{{\sdm B}\XS}
\def\CD{{\sdm C}\XS}
\def\DD{{\sdm D}\XS}
\def\ED{{\sdm E}\XS}
\def\FD{{\sdm F}\XS}
\def\GD{{\sdm G}\XS}
\def\HD{{\sdm H}\XS}
\def\ID{{\sdm I}\XS}
\def\JD{{\sdm J}\XS}
\def\KD{{\sdm K}\XS}
\def\LD{{\sdm L}\XS}
\def\MD{{\sdm M}\XS}
\def\ND{{\sdm N}\XS}
\def\OD{{\sdm O}\XS}
\def\PD{{\sdm P}\XS}
\def\QD{{\sdm Q}\XS}
\def\RD{{\sdm R}\XS}
\def\SD{{\sdm S}\XS}
\def\TD{{\sdm T}\XS}
\def\UD{{\sdm U}\XS}
\def\VD{{\sdm V}\XS}
\def\WD{{\sdm W}\XS}
\def\XD{{\sdm X}\XS}
\def\YD{{\sdm Y}\XS}
\def\ZD{{\sdm Z}\XS}

\def\ADp{{\sdmp A}\XS}
\def\BDp{{\sdmp B}\XS}
\def\CDp{{\sdmp C}\XS}
\def\DDp{{\sdmp D}\XS}
\def\EDp{{\sdmp E}\XS}
\def\FDp{{\sdmp F}\XS}
\def\GDp{{\sdmp G}\XS}
\def\HDp{{\sdmp H}\XS}
\def\IDp{{\sdmp I}\XS}
\def\JDp{{\sdmp J}\XS}
\def\KDp{{\sdmp K}\XS}
\def\LDp{{\sdmp L}\XS}
\def\MDp{{\sdmp M}\XS}
\def\NDp{{\sdmp N}\XS}
\def\ODp{{\sdmp O}\XS}
\def\PDp{{\sdmp P}\XS}
\def\QDp{{\sdmp Q}\XS}
\def\RDp{{\sdmp R}\XS}
\def\SDp{{\sdmp S}\XS}
\def\TDp{{\sdmp T}\XS}
\def\UDp{{\sdmp U}\XS}
\def\VDp{{\sdmp V}\XS}
\def\WDp{{\sdmp W}\XS}
\def\XDp{{\sdmp X}\XS}
\def\YDp{{\sdmp Y}\XS}
\def\ZDp{{\sdmp Z}\XS}

\def\ADt{\trans{\sdm A}\XS} 
\def\BDt{\trans{\sdm B}\XS}
\def\CDt{\trans{\sdm C}\XS}
\def\DDt{\trans{\sdm D}\XS}
\def\EDt{\trans{\sdm E}\XS}
\def\FDt{\trans{\sdm F}\XS}
\def\GDt{\trans{\sdm G}\XS}
\def\HDt{\trans{\sdm H}\XS}
\def\IDt{\trans{\sdm I}\XS}
\def\JDt{\trans{\sdm J}\XS}
\def\KDt{\trans{\sdm K}\XS}
\def\LDt{\trans{\sdm L}\XS}
\def\MDt{\trans{\sdm M}\XS}
\def\NDt{\trans{\sdm N}\XS}
\def\ODt{\trans{\sdm O}\XS}
\def\PDt{\trans{\sdm P}\XS}
\def\QDt{\trans{\sdm Q}\XS}
\def\RDt{\trans{\sdm R}\XS}
\def\SDt{\trans{\sdm S}\XS}
\def\TDt{\trans{\sdm T}\XS}
\def\UDt{\trans{\sdm U}\XS}
\def\VDt{\trans{\sdm V}\XS}
\def\WDt{\trans{\sdm W}\XS}
\def\XDt{\trans{\sdm X}\XS}
\def\YDt{\trans{\sdm Y}\XS}
\def\ZDt{\trans{\sdm Z}\XS}

\def\aD{{\sdm a}\XS}
\def\bD{{\sdm b}\XS}
\def\cD{{\sdm c}\XS}
\def\dD{{\sdm d}\XS}
\def\eD{{\sdm e}\XS}
\def\fD{{\sdm f}\XS}
\def\gD{{\sdm g}\XS}
\def\hD{{\sdm h}\XS}
\def\iD{{\sdm i}\XS}
\def\jD{{\sdm j}\XS}
\def\kD{{\sdm k}\XS}
\def\lD{{\sdm l}\XS}
\def\mD{{\sdm m}\XS}
\def\nD{{\sdm n}\XS}
\def\oD{{\sdm o}\XS}
\def\pD{{\sdm p}\XS}
\def\qD{{\sdm q}\XS}
\def\rD{{\sdm r}\XS}
\def\sD{{\sdm s}\XS}
\def\tD{{\sdm t}\XS}
\def\uD{{\sdm u}\XS}
\def\vD{{\sdm D}\XS}
\def\wD{{\sdm w}\XS}
\def\xD{{\sdm x}\XS}
\def\yD{{\sdm y}\XS}
\def\zD{{\sdm z}\XS}

\def\aDp{{\sdmp a}\XS}
\def\bDp{{\sdmp b}\XS}
\def\cDp{{\sdmp c}\XS}
\def\dDp{{\sdmp d}\XS}
\def\eDp{{\sdmp e}\XS}
\def\fDp{{\sdmp f}\XS}
\def\gDp{{\sdmp g}\XS}
\def\hDp{{\sdmp h}\XS}
\def\iDp{{\sdmp i}\XS}
\def\jDp{{\sdmp j}\XS}
\def\kDp{{\sdmp k}\XS}
\def\lDp{{\sdmp l}\XS}
\def\mDp{{\sdmp m}\XS}
\def\nDp{{\sdmp n}\XS}
\def\oDp{{\sdmp o}\XS}
\def\pDp{{\sdmp p}\XS}
\def\qDp{{\sdmp q}\XS}
\def\rDp{{\sdmp r}\XS}
\def\sDp{{\sdmp s}\XS}
\def\tDp{{\sdmp t}\XS}
\def\uDp{{\sdmp u}\XS}
\def\vDp{{\sdmp v}\XS}
\def\wDp{{\sdmp w}\XS}
\def\xDp{{\sdmp x}\XS}
\def\yDp{{\sdmp y}\XS}
\def\zDp{{\sdmp z}\XS}

\def\aDt{\trans{\sdm a}\XS}
\def\bDt{\trans{\sdm b}\XS}
\def\cDt{\trans{\sdm c}\XS}
\def\dDt{\trans{\sdm d}\XS}
\def\eDt{\trans{\sdm e}\XS}
\def\fDt{\trans{\sdm f}\XS}
\def\gDt{\trans{\sdm g}\XS}
\def\hDt{\trans{\sdm h}\XS}
\def\iDt{\trans{\sdm i}\XS}
\def\jDt{\trans{\sdm j}\XS}
\def\kDt{\trans{\sdm k}\XS}
\def\lDt{\trans{\sdm l}\XS}
\def\mDt{\trans{\sdm m}\XS}
\def\nDt{\trans{\sdm n}\XS}
\def\oDt{\trans{\sdm o}\XS}
\def\pDt{\trans{\sdm p}\XS}
\def\qDt{\trans{\sdm q}\XS}
\def\rDt{\trans{\sdm r}\XS}
\def\sDt{\trans{\sdm s}\XS}
\def\tDt{\trans{\sdm t}\XS}
\def\uDt{\trans{\sdm u}\XS}
\def\vDt{\trans{\sdm v}\XS}
\def\wDt{\trans{\sdm w}\XS}
\def\xDt{\trans{\sdm x}\XS}
\def\yDt{\trans{\sdm y}\XS}
\def\zDt{\trans{\sdm z}\XS}

\def\Av{{\sbv A}\XS}	\def\av{{\sbv a}\XS}
\def\Bv{{\sbv B}\XS}	\def\bv{{\sbv b}\XS}
\def\Cv{{\sbv C}\XS}	\def\cv{{\sbv c}\XS}
\def\Dv{{\sbv D}\XS}	\def\dv{{\sbv d}\XS}
\def\Ev{{\sbv E}\XS}	\def\ev{{\sbv e}\XS}
\def\Fv{{\sbv F}\XS}	\def\fv{{\sbv f}\XS}
\def\Gv{{\sbv G}\XS}	\def\gv{{\sbv g}\XS}
\def\Hv{{\sbv H}\XS}	\def\hv{{\sbv h}\XS}
\def\Iv{{\sbv I}\XS}	\def\iv{{\sbv i}\XS}
\def\Jv{{\sbv J}\XS}	\def\jv{{\sbv j}\XS}
\def\Kv{{\sbv K}\XS}	\def\kv{{\sbv k}\XS}
\def\Lv{{\sbv L}\XS}	\def\lv{{\sbv l}\XS}
\def\Mv{{\sbv M}\XS}	\def\mv{{\sbv m}\XS}
\def\Nv{{\sbv N}\XS}	\def\nv{{\sbv n}\XS}
\def\Ov{{\sbv O}\XS}	\def\ov{{\sbv o}\XS}
\def\Pv{{\sbv P}\XS}	\def\pv{{\sbv p}\XS}
\def\Qv{{\sbv Q}\XS}	\def\qv{{\sbv q}\XS}
\def\Rv{{\sbv R}\XS}	\def\rv{{\sbv r}\XS}
\def\Sv{{\sbv S}\XS}	\def\sv{{\sbv s}\XS}
\def\Tv{{\sbv T}\XS}	\def\tv{{\sbv t}\XS}
\def\Uv{{\sbv U}\XS}	\def\uv{{\sbv u}\XS}
\def\Vv{{\sbv V}\XS}	\def\vv{{\sbv v}\XS}
\def\Wv{{\sbv W}\XS}	\def\wv{{\sbv w}\XS}
\def\Xv{{\sbv X}\XS}	\def\xv{{\sbv x}\XS}
\def\Yv{{\sbv Y}\XS}	\def\yv{{\sbv y}\XS}
\def\Zv{{\sbv Z}\XS}	\def\zv{{\sbv z}\XS}

\def\Avp{{\sbvp A}\XS}	\def\avp{{\sbvp a}\XS}
\def\Bvp{{\sbvp B}\XS}	\def\bvp{{\sbvp b}\XS}
\def\Cvp{{\sbvp C}\XS}	\def\cvp{{\sbvp c}\XS}
\def\Dvp{{\sbvp D}\XS}	\def\dvp{{\sbvp d}\XS}
\def\Evp{{\sbvp E}\XS}	\def\evp{{\sbvp e}\XS}
\def\Fvp{{\sbvp F}\XS}	\def\fvp{{\sbvp f}\XS}
\def\Gvp{{\sbvp G}\XS}	\def\gvp{{\sbvp g}\XS}
\def\Hvp{{\sbvp H}\XS}	\def\hvp{{\sbvp h}\XS}
\def\Ivp{{\sbvp I}\XS}	\def\ivp{{\sbvp i}\XS}
\def\Jvp{{\sbvp J}\XS}	\def\jvp{{\sbvp j}\XS}
\def\Kvp{{\sbvp K}\XS}	\def\kvp{{\sbvp k}\XS}
\def\Lvp{{\sbvp L}\XS}	\def\lvp{{\sbvp l}\XS}
\def\Mvp{{\sbvp M}\XS}	\def\mvp{{\sbvp m}\XS}
\def\Nvp{{\sbvp N}\XS}	\def\nvp{{\sbvp n}\XS}
\def\Ovp{{\sbvp O}\XS}	\def\ovp{{\sbvp o}\XS}
\def\Pvp{{\sbvp P}\XS}	\def\pvp{{\sbvp p}\XS}
\def\Qvp{{\sbvp Q}\XS}	\def\qvp{{\sbvp q}\XS}
\def\Rvp{{\sbvp R}\XS}	\def\rvp{{\sbvp r}\XS}
\def\Svp{{\sbvp S}\XS}	\def\svp{{\sbvp s}\XS}
\def\Tvp{{\sbvp T}\XS}	\def\tvp{{\sbvp t}\XS}
\def\Uvp{{\sbvp U}\XS}	\def\uvp{{\sbvp u}\XS}
\def\Vvp{{\sbvp V}\XS}	\def\vvp{{\sbvp v}\XS}
\def\Wvp{{\sbvp W}\XS}	\def\wvp{{\sbvp w}\XS}
\def\Xvp{{\sbvp X}\XS}	\def\xvp{{\sbvp x}\XS}
\def\Yvp{{\sbvp Y}\XS}	\def\yvp{{\sbvp y}\XS}
\def\Zvp{{\sbvp Z}\XS}	\def\zvp{{\sbvp z}\XS}

\def\Avt{\trans{\sbv A}\XS}	\def\avt{\trans{\sbv a}\XS}
\def\Bvt{\trans{\sbv B}\XS}	\def\bvt{\trans{\sbv b}\XS}
\def\Cvt{\trans{\sbv C}\XS}	\def\cvt{\trans{\sbv c}\XS}
\def\Dvt{\trans{\sbv D}\XS}	\def\dvt{\trans{\sbv d}\XS}
\def\Evt{\trans{\sbv E}\XS}	\def\evt{\trans{\sbv e}\XS}
\def\Fvt{\trans{\sbv F}\XS}	\def\fvt{\trans{\sbv f}\XS}
\def\Gvt{\trans{\sbv G}\XS}	\def\gvt{\trans{\sbv g}\XS}
\def\Hvt{\trans{\sbv H}\XS}	\def\hvt{\trans{\sbv h}\XS}
\def\Ivt{\trans{\sbv I}\XS}	\def\ivt{\trans{\sbv i}\XS}
\def\Jvt{\trans{\sbv J}\XS}	\def\jvt{\trans{\sbv j}\XS}
\def\Kvt{\trans{\sbv K}\XS}	\def\kvt{\trans{\sbv k}\XS}
\def\Lvt{\trans{\sbv L}\XS}	\def\lvt{\trans{\sbv l}\XS}
\def\Mvt{\trans{\sbv M}\XS}	\def\mvt{\trans{\sbv m}\XS}
\def\Nvt{\trans{\sbv N}\XS}	\def\nvt{\trans{\sbv n}\XS}
\def\Ovt{\trans{\sbv O}\XS}	\def\ovt{\trans{\sbv o}\XS}
\def\Pvt{\trans{\sbv P}\XS}	\def\pvt{\trans{\sbv p}\XS}
\def\Qvt{\trans{\sbv Q}\XS}	\def\qvt{\trans{\sbv q}\XS}
\def\Rvt{\trans{\sbv R}\XS}	\def\rvt{\trans{\sbv r}\XS}
\def\Svt{\trans{\sbv S}\XS}	\def\svt{\trans{\sbv s}\XS}
\def\Tvt{\trans{\sbv T}\XS}	\def\tvt{\trans{\sbv t}\XS}
\def\Uvt{\trans{\sbv U}\XS}	\def\uvt{\trans{\sbv u}\XS}
\def\Vvt{\trans{\sbv V}\XS}	\def\vvt{\trans{\sbv v}\XS}
\def\Wvt{\trans{\sbv W}\XS}	\def\wvt{\trans{\sbv w}\XS}
\def\Xvt{\trans{\sbv X}\XS}	\def\xvt{\trans{\sbv x}\XS}
\def\Yvt{\trans{\sbv Y}\XS}	\def\yvt{\trans{\sbv y}\XS}
\def\Zvt{\trans{\sbv Z}\XS}	\def\zvt{\trans{\sbv z}\XS}

\def\alphab{{\sbm \alpha}\XS}		\def\Gammab{{\sbm \Gamma}\XS}
\def\betab{{\sbm \beta}\XS}		\def\Deltab{{\sbm \Delta}\XS}
\def\gammab{{\sbm \gamma}\XS} 		\def\Thetab{{\sbm \Theta}\XS}
\def\deltab{{\sbm \delta}\XS}		\def\Lambdab{{\sbm \Lambda}\XS}
\def\epsilonb{{\sbm \epsilon}\XS} 	\def\Xib{{\sbm \Xi}\XS}
\def\varepsilonb{{\sbm \varepsilon}\XS} \def\Pib{{\sbm \Pi}\XS}
\def\zetab{{\sbm \zeta}\XS}		\def\Sigmab{{\sbm \Sigma}\XS}
\def\etab{{\sbm \eta}\XS}		\def\Varsigmab{{\sbm \Varsigma}\XS}
\def\thetab{{\sbm \theta}\XS}		\def\Upsilonb{{\sbm \Upsilon}\XS}
\def\varthetab{{\sbm \vartheta}\XS}	\def\Phib{{\sbm \Phi}\XS}
\def\iotab{{\sbm \iota}\XS}		\def\Psib{{\sbm \Psi}\XS}
\def\kappab{{\sbm \kappa}\XS}		\def\Omegab{{\sbm \Omega}\XS}
\def\lambdab{{\sbm \lambda}\XS}
\def\mub{{\sbm \mu}\XS}
\def\nub{{\sbm \nu}\XS}
\def\xib{{\sbm \xi}\XS} 
\def\pib{{\sbm \pi}\XS}
\def\varpib{{\sbm \varpi}\XS}
\def\rhob{{\sbm \rho}\XS}
\def\varrhob{{\sbm \varrho}\XS}
\def\sigmab{{\sbm \sigma}\XS}
\def\varsigmab{{\sbm \varsigma}\XS}
\def\phib{{\sbm \phi}\XS}
\def\varphib{{\sbm \varphi}\XS}
\def\chib{{\sbm \chi}\XS}
\def\psib{{\sbm \psi}\XS} 
\def\omegab{{\sbm \omega}\XS} 
\def\taub{{\sbm \tau}\XS}
\def\upsilonb{{\sbm \upsilon}\XS}

\def\alphabt{\trans{\sbm \alpha}\XS}		 \def\Gammabt{\trans{\sbm \Gamma}\XS}
\def\betabt{\trans{\sbm \beta}\XS}		 \def\Deltabt{\trans{\sbm \Delta}\XS}
\def\gammabt{\trans{\sbm \gamma}\XS} 		 \def\Thetabt{\trans{\sbm \Theta}\XS}
\def\deltabt{\trans{\sbm \delta}\XS}		 \def\Lambdabt{\trans{\sbm \Lambda}\XS}
\def\epsilonbt{\trans{\sbm \epsilon}\XS} 	 \def\Xibt{\trans{\sbm \Xi}\XS}
\def\varepsilonbt{\trans{\sbm \varepsilon}\XS}	 \def\Pibt{\trans{\sbm \Pi}\XS}
\def\zetabt{\trans{\sbm \zeta}\XS}		 \def\Sigmabt{\trans{\sbm \Sigma}\XS}
\def\etabt{\trans{\sbm \eta}\XS}		 \def\Varsigmabt{\trans{\sbm \Varsigma}\XS}
\def\thetabt{\trans{\sbm \theta}\XS}		 \def\Phibt{\trans{\sbm \Phi}\XS}
\def\varthetabt{\trans{\sbm \vartheta}\XS}	 \def\Psibt{\trans{\sbm \Psi}\XS}
\def\iotabt{\trans{\sbm \iota}\XS}		 \def\Omegabt{\trans{\sbm \Omega}\XS}
\def\kappabt{\trans{\sbm \kappa}\XS}		 \def\Upsilonbt{\trans{\sbm \Upsilon}\XS}
\def\lambdabt{\trans{\sbm \lambda}\XS}
\def\mubt{\trans{\sbm \mu}\XS}
\def\nubt{\trans{\sbm \nu}\XS}
\def\xibt{\trans{\sbm \xi}\XS} 
\def\pibt{\trans{\sbm \pi}\XS}
\def\varpibt{\trans{\sbm \varpi}\XS}
\def\rhobt{\trans{\sbm \rho}\XS}
\def\varrhobt{\trans{\sbm \varrho}\XS}
\def\sigmabt{\trans{\sbm \sigma}\XS}
\def\varsigmabt{\trans{\sbm \varsigma}\XS}
\def\phibt{\trans{\sbm \phi}\XS}
\def\varphibt{\trans{\sbm \varphi}\XS}
\def\chibt{\trans{\sbm \chi}\XS}
\def\psibt{\trans{\sbm \psi}\XS} 
\def\omegabt{\trans{\sbm \omega}\XS} 
\def\taubt{\trans{\sbm \tau}\XS}
\def\upsilonbt{\trans{\sbm \upsilon}\XS}

\def\alphabp{{\sbmp \alpha}\XS} 	     \def\Gammabp{{\sbmp \Gamma}\XS}
\def\betabp{{\sbmp \beta}\XS}		     \def\Deltabp{{\sbmp \Delta}\XS}
\def\gammabp{{\sbmp \gamma}\XS} 	     \def\Thetabp{{\sbmp \Theta}\XS}
\def\deltabp{{\sbmp \delta}\XS} 	     \def\Lambdabp{{\sbmp \Lambda}\XS}
\def\epsilonbp{{\sbmp \epsilon}\XS} 	     \def\Xibp{{\sbmp \Xi}\XS}
\def\varepsilonbp{{\sbmp \varepsilon}\XS}    \def\Pibp{{\sbmp \Pi}\XS}
\def\zetabp{{\sbmp \zeta}\XS}		     \def\Sigmabp{{\sbmp \Sigma}\XS}
\def\etabp{{\sbmp \eta}\XS}		     \def\Varsigmabp{{\sbmp \Varsigma}\XS}
\def\thetabp{{\sbmp \theta}\XS} 	     \def\Phibp{{\sbmp \Phi}\XS}
\def\varthetabp{{\sbmp \vartheta}\XS}	     \def\Psibp{{\sbmp \Psi}\XS}
\def\iotabp{{\sbmp \iota}\XS}		     \def\Omegabp{{\sbmp \Omega}\XS}
\def\kappabp{{\sbmp \kappa}\XS} 	     \def\Upsilonbp{{\sbmp \Upsilon}\XS}
\def\lambdabp{{\sbmp \lambda}\XS}
\def\mubp{{\sbmp \mu}\XS}
\def\nubp{{\sbmp \nu}\XS}
\def\xibp{{\sbmp \xi}\XS} 
\def\pibp{{\sbmp \pi}\XS}
\def\varpibp{{\sbmp \varpi}\XS}
\def\rhobp{{\sbmp \rho}\XS}
\def\varrhobp{{\sbmp \varrho}\XS}
\def\sigmabp{{\sbmp \sigma}\XS}
\def\varsigmabp{{\sbmp \varsigma}\XS}
\def\phibp{{\sbmp \phi}\XS}
\def\varphibp{{\sbmp \varphi}\XS}
\def\chibp{{\sbmp \chi}\XS}
\def\psibp{{\sbmp \psi}\XS} 
\def\omegabp{{\sbmp \omega}\XS} 
\def\taubp{{\sbmp \tau}\XS}
\def\upsilonbp{{\sbmp \upsilon}\XS}

\def\zerob{{\sbm 0}\XS}   \def\zerobp{{\sbmp 0}\XS}   \def\zerobt{\trans{\sbm 0}\XS}
\def\unb{{\sbm 1}\XS}	  \def\unbp{{\sbmp 1}\XS}     \def\unbt{\trans{\sbm 1}\XS}
\def\deuxb{{\sbm 2}\XS}   \def\deuxbp{{\sbmp 2}\XS}   \def\deuxbt{\trans{\sbm 2}\XS}
\def\troisb{{\sbm 3}\XS}  \def\troisbp{{\sbmp 3}\XS}  \def\troisbt{\trans{\sbm 3}\XS}
\def\quatreb{{\sbm 4}\XS} \def\quatrebp{{\sbmp 4}\XS} \def\quatrebt{\trans{\sbm 4}\XS}
\def\cinqb{{\sbm 5}\XS}   \def\cinqbp{{\sbmp 5}\XS}   \def\cinqbt{\trans{\sbm 5}\XS}
\def\sixb{{\sbm 6}\XS}	  \def\sixbp{{\sbmp 6}\XS}    \def\sixbt{\trans{\sbm 6}\XS}
\def\septb{{\sbm 7}\XS}   \def\septbp{{\sbmp 7}\XS}   \def\septbt{\trans{\sbm 7}\XS}
\def\huitb{{\sbm 8}\XS}   \def\huitbp{{\sbmp 8}\XS}   \def\huitbt{\trans{\sbm 8}\XS}
\def\neufb{{\sbm 9}\XS}   \def\neufbp{{\sbmp 9}\XS}   \def\neufbt{\trans{\sbm 9}\XS}

\def\eC{\mbox{$\mathbb{C}$}\XS} 		\def\eCp{\mathbb{C}\XS}
\def\eE{\mbox{$\mathbb{E}$}\XS} 		\def\eEp{\mathbb{E}\XS}
\def\eN{\mbox{$\mathbb{N}$}\XS} 		\def\eNp{\mathbb{N}\XS}
\def\eR{\mbox{$\mathbb{R}$}\XS} 		\def\eRp{\mathbb{R}\XS}
\def\eZ{\mbox{$\mathbb{Z}$}\XS} 		\def\eZp{\mathbb{Z}\XS}

\def\d#1{\,\mbox{d}#1}

\def\pth#1{\left(#1\right)}
\def\acc#1{\left\{#1\right\}}
\def\cro#1{\left[#1\right]}
\def\bars#1{\left|#1\right|}
\def\norm#1{\left\|#1\right\|}

\def\ER{\mbox{I\kern-.25em R}}
\def\EC{\mbox{C\kern-.8em C}}
\def\EZ{\mbox{Z\kern-.55em Z}}
\def\EN{\mbox{N\kern-.8em N}}

\def\dpdx#1#2{\frac{\partial {#1}}{\partial {#2}}}
\def\dpdxd#1#2{\frac{\partial^2 {#1}}{\partial {#2}^2}}
\def\dpdxdy#1#2#3{\frac{{\partial^2 {#1}}}{\partial {#2} \partial {#3}}}

\def\rem#1{}

\def\cov#1{\mbox{Cov}\cro{#1}}
\def\esp#1{\mbox{E}\cro{#1}}
\def\espx#1#2{\mbox{E}_{#1}\cro{#2}}

\def\xtheta#1{\left(\frac{x_{#1}}{\theta_{#1}}\right)}
\def\xthetap#1#2{\xtheta{#1}^{\frac{#2}{\gamma_{#1}}}}

\def\xthetak{\left(\frac{x_{k}}{\theta_{k}}\right)}
\def\xthetal{\left(\frac{x_{l}}{\theta_{l}}\right)}

\def\xthetalc{\left(\frac{x_l}{\theta_l}\right)^{c_l}}
\def\xthetakc{\left(\frac{x_k}{\theta_k}\right)^{c_k}}
\def\xthetajc{\left(\frac{x_j}{\theta_j}\right)^{c_j}}

\def\Xtheta#1{\left(\frac{X_{#1}}{\theta_{#1}}\right)}
\def\Xthetap#1#2{\Xtheta{#1}^{\frac{#2}{\gamma_{#1}}}}

\def\Xthetak{\left(\frac{X_{k}}{\theta_{k}}\right)}
\def\Xthetal{\left(\frac{X_{l}}{\theta_{l}}\right)}

\def\Xthetalc{\left(\frac{X_l}{\theta_l}\right)^{c_l}}
\def\Xthetakc{\left(\frac{X_k}{\theta_k}\right)^{c_k}}
\def\Xthetajc{\left(\frac{X_j}{\theta_j}\right)^{c_j}}

\def\xthetalc{\left(\frac{x_l}{\theta_l}\right)^{c_l}}
\def\xthetakc{\left(\frac{x_k}{\theta_k}\right)^{c_k}}
\def\xthetajc{\left(\frac{x_j}{\theta_j}\right)^{c_j}}

\def\Xthetalc{\left(\frac{X_l}{\theta_l}\right)^{c_l}}
\def\Xthetakc{\left(\frac{X_k}{\theta_k}\right)^{c_k}}
\def\Xthetajc{\left(\frac{X_j}{\theta_j}\right)^{c_j}}

\def\ppj{\left(1+\sum_{j=1}^{n} \xthetap{j}{1}\right)}
\def\ppjc{\left(1+\sum_{j=1}^{n} \xthetajc^{c_j}\right)}

\def\ppjX{\left(1+\sum_{j=1}^{n} \Xthetap{j}{1}\right)}
\def\ppjcX{\left(1+\sum_{j=1}^{n} \Xthetajc^{c_j}\right)}

\def\fx{f_{n}(\xb)}
\def\fX{f_{n}(\Xb)}

\def\fcl{\frac{\xthetap{l}{1}}{\ppj}}
\def\fck{\frac{\xthetap{k}{1}}{\ppj}}
\def\fclc{\frac{\xthetalc}{\ppjc}}

\def\fclX{\frac{\Xthetap{l}{1}}{\ppjX}}
\def\fckX{\frac{\Xthetap{k}{1}}{\ppjX}}
\def\fclcX{\frac{\Xthetalc}{\ppjcX}}

\def\fcllog{\frac{\xthetap{l}{1}\ln\xthetal}{\ppj}}
\def\fcllogd{\frac{\xthetap{l}{1}\ln^2\xthetal}{\ppj}}

\def\fcllogX{\frac{\Xthetap{l}{1}\ln\Xthetal}{\ppjX}}
\def\fcllogdX{\frac{\Xthetap{l}{1}\ln^2\Xthetal}{\ppjX}}

\def\fcd{\frac{\xthetap{l}{2}}{\ppj^2}}
\def\ktan{\quad k=1, \cdots,n}
\def\ltan{\quad l=1, \cdots,n}

\def\psil{\Psi_{r_l}(\alpha)}
\def\psik{\Psi_{r_k}(\alpha)}
\def\psikl{\Psi_{r_l r_k}(\alpha)}

\def\xmut{\pth{\frac{x_l-\mu_l}{\theta_l}}}
\def\xmutl{\pth{\frac{x_l-\mu_l}{\theta_l}}^{m_l}}
\def\xmutk{\pth{\frac{x_k-\mu_k}{\theta_k}}^{m_k}}

\def\Xmut{\pth{\frac{X_l-\mu_l}{\theta_l}}}
\def\Xmutl{\pth{\frac{X_l-\mu_l}{\theta_l}}^{m_l}}
\def\Xmutk{\pth{\frac{X_k-\mu_k}{\theta_k}}^{m_k}}

\def\fxx{f_{X_l,X_k}(x_l,x_k)}
\def\dxx{\d{x_l}\d{x_k}}

\def\cktk{\frac{c_k}{\theta_k}}
\def\cltl{\frac{c_l}{\theta_l}}

\def\mata{\left[
\barr{ccc}
\barr{c} I(\theta_l) \\ I(\theta_l,\gamma_l) \\
I(\gamma_l,\alpha) \earr
& \barr{c} I(\theta_l,\gamma_l) \\
I(\gamma_l) \\ I(\gamma_l,\alpha) \earr&
\barr{c}
I(\theta_l,\alpha)
\\ I(\gamma_l,\alpha) \\ I(\alpha) \earr \earr\right]}

\def\matb{\left[
\barr{ccc} \barr{c} 1 \\ 0 \\ 0 \earr & \barr{c} 0 \\ \gamma^2 \\
0 \earr & \barr{c} 0
\\ 0 \\ 1 \earr \earr\right]}

\def\matbc{\left[
\barr{ccc} \barr{c} a \\ b \\ c \earr & \barr{c} d \\ e \\
f \earr & \barr{c} g
\\ h \\ i \earr \earr\right]}

\title{Entropy, Information Matrix and order statistics of Multivariate Pareto, Burr and related distributions}

\author{Gholamhossein Yari$~^{1,2}$ \\ 
$~^1$ Iran University of Science and Technology, Narmak, Tehran 16844, Iran.
\\[12pt] 
Ali Mohammad-Djafari$~^2$\\ 
$~^2$ Laboratoire des Signaux et Syst\`emes ({\sc Cnrs,Sup\'elec,Ups)}, \\ 
Sup\'elec, Plateau de Moulon, 3 rue Joliot-Curie, 91192 Gif-sur-Yvette, France.}

\begin{document}
\maketitle 

\begin{abstract}
In this paper we derive the exact analytical
expressions for the information and covariance matrices of the multivariate
Burr and related distributions. These distributions arise
as tractable parametric models in reliability, actuarial science,
economics, finance and telecommunications. We show that all the
calculations can be obtained from one main moment multi
dimensional integral whose expression is obtained through some
particular change of variables. 
\\ ~\\ 
{\bf keywords:} 
Gamma and Beta functions;  Polygamma functions ;  
Information matrix;  Covariance matrix;  Multivariate Burr models.
\end{abstract}

\newpage
\section{Introduction}
\label{s1}

In this paper the exact form of Fisher information matrix for
 multivariate Pareto (IV) and related distributions is determined. It is
 well-known that the information matrix is a valuable tool for
 derivation of covariance matrix in the asymptotic distribution of maximum
 likelihood estimations (MLE). In the univariate case of the
 above distributions, the Fisher information matrix is found by
Brazauskas~\cite{Brazauskas2003}.
 As discussed in Serfling~\cite{Serfling1980}, section~\ref{s4}, under suitable regularity
 conditions, the determinant of the asymptotic covariance matrix of
 (MLE) reaches an optimal lower bound for the volume of the spread
 ellipsoid of joint estimators. In the univariate case of the
 Pareto (IV), this optimality property of (MLE) is widely used in the
 robustness versus efficiency studies as a quantitative benchmark
 for efficiency considerations (Brazauskas and
Serfling~\cite{Brazauskas2000a, Brazauskas2000b}, Brazauskas~\cite{Brazauskas2002},
 Hampel et al~\cite{Hampel1986}, 
Huber~\cite{Huber1981},  Klugman~\cite{Klugman1998},
Kimber~\cite{Kimber1983a,Kimber1983b} and
 Lehmann~\cite{Lehmann1983}, Chapter 5). These
distributions are suitable for situations involving relatively high probability
 in the upper tails. More specifically, such models
 have been formulated in the context of actuarial
 science, reliability, economics, finance and teletrafic. These models
 arise whenever we need to infer the
 distributions of variables such as sizes of insurance
 claims, sizes of firms, income in a population of people, stock
 price fluctuations and length of telephone calls. For a broad discussion of
 Pareto models and diverse applications see Arnold~\cite{Arnold1983},
 Johnson, Kotz and Balakrishnan~\cite{Johnson1994}, Chapter 19.
Gomes, Selman and Crato~\cite{Gomes1997} have recently discovered
 Pareto (IV) tail behavior in the cost distributions of
 combinatorial search algorithms.
\rem{The application of Burr XII
 distribution in univariate case for lifetime data.
 This distribution provide an option in the analysis of lifetime
 data; published applications include the analysis of business
 failure data,
 , the efficiency of analgesics in  clinical trials(Wingo~\cite{Wingo}) and
the failure time of electronic
 components (Wang, Keats and Zimmer~\cite{Zimmer}).
}

This paper is organized as follows: 
Multivariate Pareto and Burr distribution are introduced and presented in section~\ref{s2}.
Elements of the information and covariance matrix for multivariate Pareto
(IV) distribution is derived in section~\ref{s3}.
 Elements of the information matrices for
 Multivariate Burr, Pareto (III), and Pareto (II) distributions
are derived in section~\ref{s4}.
 Conclusion is presented in section~\ref{conclusion}.
 Derivation of first and second derivatives of the log density and the main moment
  integral calculation are given
 in Appendices $A$ and $B$ .

\section{Multivariate Pareto and Burr distributions}
\label{s2}

As discussed in Arnold~\cite{Arnold1983} Chapter 3, a hierarchy of Pareto distribution
is established
by starting with the classical Pareto (I) distribution and subsequently
 introducing additional
parameters related to location, scale, shape and inequality (Gini index).
Such an approach leads to a very general family of distributions,  called
the Pareto (IV)
family, with the cumulative distribution function
\beq
 \label{eq1}
F_{X}(x)=1-\pth{1+(\frac{x-\mu}{\theta})^\frac{1}{\gamma}}^{-\alpha}, \quad x> \mu,
\eeq
where $-\infty<\mu<+\infty$ is the location parameter,  $\theta>0$ is the scale
parameter, $\gamma>0$
is the inequality parameter and $\alpha>0$ is the shape parameter which
 characterizes
 the tail of the distribution.
 We note this distribution by Pareto (IV) $(\mu, \theta, \gamma, \alpha)$.
 Parameter $\gamma$ is called the inequality parameter
  because of its
 interpretation in the economics context. That is, if we choose $\alpha=1$
 and $\mu=0$ in
 expression (\ref{eq1}),  the parameter $(\gamma\leq 1)$ is precisely the
 Gini index of inequality.  For the Pareto  (IV) $(\mu, \theta, \gamma, \alpha)$
 distribution,  we have the density function

\beq
 \label{eq2}
f_{X}(x)=\frac{\alpha\pth{\frac{x-\mu}{\theta}}^{\frac{1}{\gamma}-1}}
{\theta\gamma\pth{1+(\frac{x-\mu}{\theta})^\frac{1}{\gamma}}^{\alpha+1}},  \quad x> \mu.
\eeq

The density of the $n$-dimensional Pareto (IV) distribution is

\beq
\label{eq3}
f_{n}(\xb)=\pth{1+\sum_{j=1}^n(\frac{x_{j}-\mu_{j}}{\theta_{j}})^{\frac{1}{\gamma_j}}}^
{-(\alpha+n)}\prod_{i=1}^n \frac{\alpha+i-1}{\theta_i\gamma_{i}}
(\frac{x_{i}-\mu_{i}}{\theta_{i}})^{\frac{1}{\gamma_i}-1},
\quad x_{i}>\mu_{i},
\eeq
where $\xb=[x_1,\cdots,x_n]$, $x_{i}>\mu_{i}$,   $\alpha>0$,   $\gamma_{i}>0$,   $\theta_{i}>0$ for $i=1,  \cdots, n$.
One of the main properties of this distribution is that,
the joint density of any subset of the components of a Pareto random
vector is again of the form $(\ref{eq3})$~\cite{Arnold1983}.

The $n$-dimensional Burr distribution has the density
\beq
 \label{eq4}
f_{n}(\xb)=\pth{1+\sum_{j=1}^n(
\frac{x_{j}-\mu_{j}}{\theta_{j}})^{c_{j}}}^{-(\alpha+n)}\prod_{i=1}^n \frac{(\alpha+i-1)c_{i}}{\theta_i}
(\frac{x_{i}-\mu_{i}}{\theta_{i}})^{{c_i}-1},
\quad x_{i}>\mu_{i},
\eeq
where $x_{i}>\mu_{i}$,   $\alpha>0$, $c_{i}>0$,   $\theta_{i}>0$ for $i=1,  \cdots, n$.
We note that the multivariate Burr distribution is equivalent to the multivariate Pareto
distribution
with $\frac{1}{\gamma_i}=c_i$.

\section{Information Matrix for Multivariate Pareto (IV)}
\label{s3}

Suppose $X$ is a random vector with the probability density function $f_\Theta(. )$ where
 $\Theta=(\theta_{1}, \theta_{2}, . . . , \theta_{K})$.
 The information matrix $I(\Theta)$ is the $K\times K$ matrix
 with elements
\beq
I_{ij}(\Theta)=-\espx{\Theta}{\dpdxdy{\ln f_{\Theta}(\Xb)}{\theta_i}{\theta_j}},
\quad i,j=1,\cdots K.
\eeq
 For the multivariate Pareto (IV), we have $\Theta = (\mu_1, ..., \mu_n, \theta_1,...,
 \theta_n, \gamma_1, ..., \gamma_n, \alpha)$.
 In order to make the multivariate Pareto (IV) distribution a regular family
 (in terms of maximum likelihood estimation), we assume that $\mu$ is known and,
 without loss of generality, equal to 0. In this case information matrix is $(2n+1)\times(2n+1)$.
 Thus, further treatment is based on the following multivariate density function

\beq
\label{chegali}
\fx=\ppj^{-(\alpha+n)}\prod_{i=1}^n \frac{\alpha+i-1}{\theta_i\gamma_{i}}
\xtheta{i}^{\frac{1}{\gamma_i}-1}
,\quad x_{i} >0.
\eeq
The log-density is:
\beqn
\label{eq5}
\ln f_{n}(\xb) &=&
\sum_{i=1}^n\cro{ \ln (\alpha+i-1) -
 \ln \theta_i +
\pth{\frac{1}{\gamma_i}-1} \ln \pth{\frac{x_i}{\theta_i}}- \ln \gamma_i}
\nonumber \\
&&
-(\alpha+n) \ln
\ppj.
\eeqn
Since the information matrix $I(\Theta)$ is symmetric it is enough to find elements
  $I_{ij}(\Theta)$, where $1 \leq i \leq j \leq  2n+1$.
The required first and second partial derivatives of the above expression
are given in the Appendix $A$.
Looking at these expressions, we see that to determine the expression of the
 information matrix and score functions,  we need to find
  the expressions of:
\[
 \esp{\ln {\ppjX}}, \quad \esp{\Xthetap{l}{r_l}}, \quad \esp{\Xthetap{l}{l}},
\]
\[
\esp{\ln\Xtheta{l}}, \quad \esp{{\Xthetap{l}{r_l}}\ln\Xtheta{l}}, \quad\esp{\fclX},
\]
and the general terms
\[
{\esp{\frac{ {\Xthetap{l}{n_{1}}} {\Xthetap{k}{n_{2}}}
\ln^{n_{4}}\Xthetak\ln^{n_{3}}\Xthetal}{\ppjX^{n_{5}}}}}, 
\quad (n_{1}, n_{2}>-1)\in\ER,
n_{3}, n_{4} \in\EN^+ \mbox{~and~} n_{5}\in\ER^+.
\]

\subsection{Main strategy to obtain expressions of the expectations}
~
Derivation of these  expressions are based on the following strategy:
first, we derive an analytical expression for the
following integral
\beq
\label{e1}
\esp{\prod_{i=1}^n \Xthetap{i}{r_i}}=
\int_{0}^{+\infty} \cdots \int_{0}^{+\infty}\prod_{i=1}^n \xthetap{i}{r_i}f_{n}(\xb)\d{\xb},
\eeq
and then, we show that all the other expressions can be found easily from it.
We consider this derivation as one of the main contributions of this work.
This derivation is given in the Appendix $B$.
The result is the following:
\beqn
\label{e2}
\esp{\prod_{i=1}^n \Xthetap{i}{r_i}}&=&
\int_{0}^{+\infty} \cdots \int_{0}^{+\infty}\prod_{i=1}^n \xthetap{i}{r_i}f_{n}(\xb)\d{\xb}=
\nonumber \\
&&
 \frac{\Gamma(\alpha-\sum_{i=1}^n r_{i}) \prod_{i=1}^n
  \Gamma(r_{i}+1)}{\Gamma(\alpha)},
\nonumber \\
&& \sum_{i=1}^n r_{i}< \alpha,\quad r_{i}>-1,  r_{i}\in\ER , 
\eeqn
where $\Gamma$ is the usual Gamma function,
\[
\Gamma_{r_l r_k}\left(\alpha-\sum_{i=1}^n r_{i}\right)=
\dpdxdy{\Gamma\left(\alpha-\sum_{i=1}^n r_{i}\right)}{r_k}{r_l},
\quad 1\leq {l, k} \leq n,
\]
\[
\Psi^{(n)} (z)=\frac{d^n}{d z^n} \pth{\frac{\Gamma'(z)}{\Gamma(z)}},\quad z>0,\quad\quad
\frac{\partial^{(m+n)}}{\partial r_l^{m}\partial r_k^{n}}
 \pth{\frac{\Gamma_{r_l r_k}(z)}{\Gamma(z)}}=
\Psi^{(m+n)}(z),\quad z >0
\]
and integers $n, m\geq 0$ (Abramowitz and Stegun~\cite{Abramowitz1972}).
 Specifically, we use digamma $\Psi(z) = \Psi^{(.)}(z)$, trigamma $\Psi'(z)$ and
 $\Psi_{r_l r_k}(z)$ functions.
To confirm the regularity of $\ln f_{n}(\xb)$ and evaluation the
expected Fisher information matrix, we take expectations of first and second order
 partial derivatives of (\ref{eq5}).
All the other expressions can be derived from this main result.
Taking of derivative with respect to
$\alpha$, from the both sides of the relation
\[
1=\int_{0}^{+\infty} f_{n}(\xb) \d{\xb},
\]
leads to
\beq
\esp{\ln {\ppjX}}=\sum_{i=1}^n \frac{1}{\alpha+i-1}.
\eeq
From relation $(\ref{e2})$, for a pair of $(l, k)$ we have
\beq
\label{e3}
\varphi(r_l,r_k)=\esp{\Xthetap{l}{r_l}\Xthetap{k}{r_k}}=
\frac{\Gamma(\alpha-r_l-r_k)\Gamma(r_l+1)\Gamma(r_k+1)}{\Gamma(\alpha)},
\eeq
and
\beq
\label{e4}
\frac{\partial^{(n_3+n_4)}}{\partial r_l^{n_3}\partial r_k^{n_4}}\varphi(r_l=n_{1},r_k=n_{2})=
\esp{{\Xthetap{l}{n_{1}}} {\Xthetap{k}{n_{2}}
\ln^{n_{4}}\Xthetak\ln^{n_{3}}\Xthetal}}.
\eeq
From relation $(\ref{e3})$, ~at $r_k=0$ we obtain
\beq
\label{e5}
\esp{\Xthetap{l}{r_l}}=
\frac{\Gamma(\alpha-r_{l})\Gamma(r_{l}+1)}{\Gamma(\alpha)},
\eeq
 and evaluating this expectation at $r_{l}=1$, we obtain
 \beq
\label{e6}
\esp{\Xthetap{l}{l}}=\frac{1}{\alpha-1}.
 \eeq
Writing the expression of the expectation
\[
\esp{\fclX}
\]
as $\mbox{~E}_{\alpha}$ to emphasis the role of the parameter
  $\alpha$ in $(\ref{chegali})$, it can easily be shown that
\beq
\espx{\alpha}{\fclX}=
  \frac{\alpha}{\alpha+n}\espx{\alpha+1}{\Xthetap{l}{1}}.
  \eeq
Using $(\ref{e6})$ with $\alpha$ replaced by $\alpha+1$, we now obtain
  an expression for the last expectation as
\[
\espx{\alpha}{\fclX}=\frac{1}{\alpha+n}.
\]
Differentiating $(\ref{e5})$ with respect to $r_{l}$,
and replacing for $r_{l}=0$ and $r_{l}=1$, we obtain the following relations:
\beq
\esp{\ln \Xtheta{l}}=\gamma_{l}\cro{{\Gamma'(1)-\Psi(\alpha)}},
\eeq
\beq
 \esp{\Xthetap{l}{1}\ln \Xtheta{l}}
 = \gamma_{l}\cro{\frac{\Gamma'(2)-\Psi(\alpha-1)}{\alpha-1}},
 \eeq
and
 \beq
 \espx{\alpha}{\fcllogX}=
\frac{\alpha}{\alpha+n}
\espx{\alpha+1}{\Xthetap{l}{1}\ln\Xtheta{l}}.
\eeq

\subsection{Expectations of the score functions} ~
The expectations of the first three partial derivations of the first order follow immediately from the
 corresponding results for their three corresponding parameters and we obtain:
\[
\esp{\dpdx{\ln\fX}{\alpha}}=\sum_{i=1}^n \frac{1}{\alpha+i-1}
 -\esp{\ln \ppjX}=0,
\]
\[
\esp{\dpdx{\ln \fX}{\theta_l}} = -\frac{1}{\theta_{l}\gamma_{l}} +
\pth{\frac{\alpha+n}{\theta_{l} \gamma_{l}}} \esp{\fclX}=0,
\]
\[
\esp{\dpdx{\ln\fX}{\gamma_l}}=
-\frac{1}{\gamma_l}-\frac{1}{\gamma_l^2}
\esp{\ln\Xtheta{l}}
+\pth{\frac{\alpha+n}{\gamma_l^2}}
\esp{\fcllogX}=0.
\]

\subsection{The expected Fisher information matrix}
Main strategy is again based d on the integral $(\ref{e2})$ which is presented in the Appendix $B$.
However, derivation of the following expressions can be obtained mecanically
but after some tedious algebraic simplifications :
\beq
I_{\xb}(\alpha)=
\sum_{i=1}^n \frac{1}{\pth{\alpha+i-1}^2},
\eeq
\beq
I_{\xb}(\theta_l,\alpha)=-\frac{1}{\theta_l\gamma_l\pth{\alpha+n}},
\eeq
\beq
I_{\xb}(\gamma_l,\alpha)=-\frac{1}{\gamma_l\pth{\alpha+n}}\cro{\Gamma'(2)-\Psi(\alpha)}, \quad l=1,\cdots,n,
\eeq
\beq
I_{\xb}(\theta_l)=\frac{\alpha+n-1}{\theta_l^2\gamma_l^2\pth{\alpha+n+1}}, \quad l=1,\cdots,n,
\eeq
\beqn
I_{\xb}(\gamma_l)&=&
\frac{\alpha+n-1}{\gamma_l^2(\alpha+n+1)}
\cro{\frac{\Gamma''(\alpha)}{\Gamma(\alpha)}+\Gamma''(1)+1}
\nonumber \\
&&
+\frac{2(\alpha+n-2)}{\gamma_l^2(\alpha+n+1)}\cro{\Gamma'(1)-\Psi(\alpha)} \nonumber \\
&&
-\frac{2(\alpha+n-1)}{\gamma_l^2(\alpha+n+1)}\cro{\Gamma'(1)\Psi(\alpha)} , \quad l=1,\cdots,n,
\eeqn
\beq
I_{\xb}(\theta_l,\theta_k)=-\frac{1}{\theta_l\gamma_l\gamma_k\theta_k \pth{\alpha+n+1}},
\quad k\neq l,
\eeq
\beq
I_{\xb}(\gamma_l,\gamma_k)=
\frac{-1}{\gamma_l\gamma_k \pth{\alpha+n+1}}
\cro{\pth{\Gamma'(2)}^2-\Gamma'(2)\pth{ \Psi_{r_l}(\alpha)+
\Psi_{r_k} (\alpha))+\Psi_{r_l}{_{r_k}}(\alpha)}}, \,  
k\neq l,
\eeq
\beq
I_{\xb}(\theta_l,\gamma_k)=-\frac{1}{\theta_l\gamma_l\gamma_k \pth{\alpha+n+1}}
\cro{\Gamma'(2)-\Psi_{r_k}  (\alpha)},
\quad k\neq l, \qquad 
\eeq
\beq
I_{\xb}(\theta_l,\gamma_l)=\frac{\alpha+n-1}{\theta_l\gamma_l^2 \pth{\alpha+n+1}}
\cro{\Gamma'(2)-\Psi(\alpha)}-\cro{\frac{1}{\theta_l\gamma_l^2 \pth{\alpha+n+1}}},
\quad l=1,\cdots,n.
\eeq
Thus the information matrix, $I_{\mbox{~MP(IV)}}(\Theta)$, for the
multivariate Pareto (IV) $ (0, \theta, \gamma, \alpha)$ distribution is
\beq
I_{\mbox{~MP(IV)}}(\Theta)=
\left[
\barr{ccc}
  \barr{c} I(\theta_l, \theta_k) \\ I(\theta_l, \gamma_k) \\ I(\theta_l, \alpha) \earr
& \barr{c} I(\theta_l, \gamma_k) \\  I(\gamma_l, \gamma_k) \\ I(\gamma_l, \alpha)   \earr
& \barr{c} I(\theta_l, \alpha) \\ I(\gamma_l, \alpha) \\  I(\alpha)  \earr
\earr
\right].
\eeq

\subsection{Covariance matrix for multivariate Pareto (IV)}~
Since the joint density of any subset of the components of a Pareto (IV)
 random vector is again a multivariate Pareto (IV), Arnold~\cite{Arnold1983}, we can
 calculate the expectation
 \beqn
 &&\esp{\Xmutl\Xmutk}= \nonumber \\
 && \int_{0}^{\infty} \int_{0}^{\infty} {\xmutl \xmutk} \fxx \dxx= \nonumber \\
 && \frac{\Gamma(\alpha-m_{l}\gamma_{l}-m_{k}\gamma_{k})
 \Gamma(m_{l}\gamma_{l}+1)\Gamma(m_{k}\gamma_{k}+1)}{\Gamma(\alpha)},  \nonumber
 \\
 && 
m_{l}, m_{k}\in\ER,\quad  m_{l}\gamma_{l},   m_{k}\gamma_{k}>-1,\quad
  \alpha-m_{l}\gamma_{l}-m_{k}\gamma_{k}>0. 
\eeqn
Evaluating this expectation at ($m_{l}=1$, $m_{k}=0$), ($m_{l}=0$, $m_{k}=1$)
 and ($m_{l}=1$, $m_{k}=1$), we obtain
\beq
\esp{X_{l}}=\mu_{l}+\frac{\theta_{l}}{\Gamma(\alpha)}[\Gamma(\alpha-\gamma_{l})
\Gamma(\gamma_{l}+1)],\quad \gamma_{l}< \alpha,\quad\gamma_{l}>-1,
\eeq
\beq
\esp{X_{k}}=\mu_{k}+\frac{\theta_{k}}{\Gamma(\alpha)}
[\Gamma(\alpha-\gamma_{k})\Gamma(\gamma_{k}+1)],\quad\gamma_{k}< \alpha,\quad\gamma_{k}>-1,
\eeq
\beqn
\esp{X_{l}X_{k}}&=&\mu_{k}\esp{X_{l}}+\mu_{l}\esp{X_{k}}-\mu_{l}\mu_{k}  \nonumber
\\
&&
+\frac{\theta_{l}\theta_{k}}{\Gamma(\alpha)}[\Gamma(\alpha-\gamma_{l}-\gamma_{k})
\Gamma(\gamma_{l}+1)\Gamma(\gamma_{k}+1)], \quad \gamma_{l}+\gamma_{k}< \alpha,
\eeqn
\beq
\esp{X_{l}^m}=\frac{\theta_{l}^m}{\Gamma(\alpha)}[\Gamma(\alpha-m_{l}\gamma_{l})
\Gamma( m_{l}\gamma_{l}+1)],\quad \gamma_{l}m_{l} < \alpha,
\eeq
\beq
\sigma^2_{X_l}=\frac{\theta_{l}^2}{\Gamma^2(\alpha)}
\cro{\Gamma(\alpha-2\gamma_l)
\Gamma(2\gamma_l+1)\Gamma(\alpha)
-\Gamma^2(\gamma_l+1)\Gamma^2(\alpha-\gamma_l)}, \quad 2\gamma_{l}< \alpha,
\eeq
\beqn
\cov{X,Y}
&=&\frac{\theta_{l}\theta_{k}\Gamma(\gamma_{l}+1)
\Gamma(\gamma_{k}+1)}{\Gamma^2(\alpha)}[\Gamma(\alpha-\gamma_{l}-
\gamma_{k})\Gamma(\alpha)
\nonumber \\
&&
-\Gamma(\alpha-\gamma_{k})\Gamma(\alpha-\gamma_{l})] ,\nonumber
\\
&&
1\leq l\leq k\leq n ,\quad k=2,\cdots,n. 
\eeqn

\section{Special Cases}
\label{s4}

\subsection{Burr$(\theta, \gamma, \alpha)$ distribution}
The Burr family of distributions is also sufficiently flexible and enjoy
long popularity in the actuarial science literature (Daykin, {Pentik\"ainen},
 and Pesonen~\cite{Daykin1994}  and Klugman, Panjer, and
 Willmot ~\cite{Klugman1998}).
 However, this family can be treated as a special case of Pareto (IV):
Burr $(\theta, \gamma, \alpha)$ = Pareto (IV) $(0, \theta, \frac{1}{\gamma}, \alpha)$
(Klugman, Panjer, and Willmot ~\cite{Klugman1998}, p. $574$).

Since the Burr distribution is a reparametrization of Pareto (IV) $(0, \theta, \gamma, \alpha)$,
 it follows from Lehmann (8), Section 2.7,  that its information matrix
  $I_{\mbox{~B}}(\Theta)$ can
 be derived from $I_{\mbox{~P(IV)}}(\Theta)$ by $JI_{\mbox{~P(IV)}}(\Theta)J'$, where $J$ is the Jacobian matrix of
  the transformation of variables.
  Thus, the information matrix of multivariate Burr distribution,
  $I_{\mbox{~MB}}(\Theta)$ is
  then given by $JI_{\mbox{~MP(IV)}}(\Theta)J'$, where
\beq
J=
\left[
\barr{ccc}
  \barr{c}I \\ 0 \\ 1 \earr
& \barr{c} 0 \\ I\gamma^2\\ 0 \earr
& \barr{c} 1 \\ 0 \\ 1 \earr
\earr
\right]
\eeq
which is obtained by noting that $J$ is the Jacobian matrix of the transformation
$(\theta, \gamma, \alpha)\rightarrow (\theta,  \frac{1}{\gamma}, \alpha)$.
\subsection{Pareto (III) $(0, \theta, \gamma)$ distribution} ~
This is a special case of Pareto (IV) with $\alpha=1$. Therefore, last row and last column of
$I_{\mbox{~MP(IV)}}(\Theta)$ vanish (these represent
information about parameter $\alpha)$ and we obtain
\beq
I_{\mbox{~MP(III)}}(\Theta)=
\left[
\barr{cc}
  \barr{c} I(\theta_l, \theta_k) \\  I(\theta_l, \gamma_k)  \earr
& \barr{c} I(\theta_l, \gamma_k) \\ I(\gamma_l, \gamma_k)   \earr
\earr
\right],
\eeq

where we have to substitute $\alpha=1$ in all the remaining expressions.
\subsection{Pareto (II) $(0, \theta, \alpha)$ distribution} ~
This is a special case of Pareto (IV) with $\gamma=1$.
Therefore $I(\theta_l, \gamma_k)$, $I(\gamma_l, \gamma_k)$ and
$I(\gamma_l, \alpha)$ in
 $I_{\mbox{~MP(IV)}}(\Theta)$ vanish and we obtain
\beq
I_{\mbox{~MP(II)}}(\Theta)=
\left[
\barr{ccc}
  \barr{c} I(\theta_l, \theta_k)  \\ I(\theta_l, \alpha) \earr
& \barr{c}  I(\theta_l, \alpha)  \\ I(\alpha)  \earr
\earr
\right],
\eeq
where we have to substitute $\gamma=1$ in all the remaining expressions.


\section{Conclusion}
\label{conclusion}
In this paper we obtained the exact form of Fisher information and covariance matrix
for multivariate Pareto (IV) distribution.
We showed that all the calculations can be obtained from one main moment
multi dimensional integral which has been considered and whose expression is
obtained through some particular
change of variables.
A short method of obtaining some of the expectations as a function of $\alpha$ is used.
To confirm the regularity of the
$\ln f_{n}(\xb)$, we showed that
the expectations of the score functions are equal to $0$.
Information matrices of
 multivariate Burr, Pareto (III) and Pareto (II) distributions are derived as
 special cases of multivariate Pareto (IV) distribution.

\newpage
\appendix
\section{Expressions of the derivatives}
\label{appendixA}
\setcounter{equation}{0}~
In this Appendix, we give detailed expressions of all the first and second
derivatives of $\ln f_n(x)$ which are needed for obtaining the expression of
the information matrix:

\beq
\dpdx{\ln \fx}{\alpha}=
\sum_{i=1}^n \frac{1}{\alpha+i-1} - \ln{\ppj},
\eeq

\beq
\dpdx{\ln \fx}{\theta_l}=
-\frac{1}{\theta_l\gamma_l}
+\pth{\frac{\alpha+n}{\theta_l\gamma_l}}
\fcl, \quad l=1,\cdots,n,
\eeq

\beq
\dpdx{\ln \fx}{\gamma_l}=
-\frac{1}{\gamma_l}
-\frac{1}{\gamma_l^2}\ln\xtheta{l}
+\pth{\frac{\alpha+n}{\gamma_l^2}}
\frac{\xthetap{l}{1}\ln \xtheta{l}}{\ppj}, \quad l=1,\cdots,n,
\eeq

\beq
   \dpdxd{\ln \fx}{\alpha}=-\sum_{i=1}^n \frac{1}{(\alpha+i-1)^2},
\eeq

\beq
   \dpdxdy{\ln \fx}{\theta_k}{\alpha}=
   \pth{\frac{1}{\theta_k\gamma_k}}
   \fck, \ktan,
\eeq

\beq
\dpdxdy{\ln \fx}{\gamma_k}{\alpha}=
\pth{\frac{1}{\gamma_k^2}}
\frac{\xthetap{k}{1} \ln\xthetak}{\ppj}, \ktan,
\eeq

\beqn
\dpdxd{\ln \fx}{\theta_l}&=&
\frac{1}{\theta_l^2\gamma_l}
-\pth{\frac{\alpha+n}{\theta_l^2\gamma_l}}
\pth{1+\frac{1}{\gamma_l}}
\fcl
\nonumber \\
&&
+\pth{\frac{\alpha+n}{\theta_l^2\gamma_l^2}}\fcd , \ltan, 
\eeqn

\beqn
\dpdxd{\ln \fx}{\gamma_l}&=&
\frac{1}{\gamma_l^2}
+\frac{2}{\gamma^3} \ln\xthetal
-2\pth{\frac{\alpha+n}{\gamma_l^3}} \fcllog 
\nonumber \\
&&
-\pth{\frac{\alpha+n}{\gamma_l^4}} \fcllogd   
\nonumber \\
&&
+\pth{\frac{\alpha+n}{\gamma_l^4}}
\pth{\fcllog}^2, 
\ltan,
\eeqn

\beq
\dpdxdy{\ln \fx}{\theta_k}{\theta_l}=
\pth{\frac{\alpha+n}{\gamma_k\gamma_k\theta_l\theta_k}}
\frac{\xthetap{l}{1}\xthetap{k}{1}}{\ppj^2}, \quad k\neq l,
\eeq
\beq
\dpdxdy{\ln \fx}{\gamma_k}{\gamma_l}=
\pth{\frac{\alpha+n}{\gamma_l^2\gamma_k^2}}
\frac{\xthetap{l}{1}\xthetap{k}{1}\ln\xthetal\ln\xthetak}{\ppj^2}, \quad k\neq l,
\eeq

\beq
\dpdxdy{\ln \fx}{\gamma_k}{\theta_l}=
  \pth{\frac{\alpha+n}{\gamma_l\theta_l\gamma_k^2}}
 \frac{\xthetap{l}{1}\xthetap{k}{1}\ln\xthetak}{\ppj^2}, \quad k\neq l,
\eeq

\beqn
 \dpdxdy{\ln \fx}{\gamma_k}{\theta_l}
&=& \frac{1}{\theta_l\gamma_l^2}
    -\pth{\frac{\alpha+n}{\theta_l\gamma_l^2}} \fcl
 \nonumber \\
&& -\pth{\frac{\alpha+n}{\theta_l\gamma_l^3}} \fcllog 
\nonumber \\
 && +\pth{\frac{\alpha+n}{\theta_{l}\gamma_l^3}} {\pth\fcl^2} \ln\xthetal, 
\quad k\neq l.
\eeqn

\newpage
\section{Expression of the main integral}
\label{appendixB}
\setcounter{equation}{0}

This Appendix gives one of the main results of this paper which is the derivation of the expression
of the following integral
\beq
\esp{\prod_{i=1}^n \Xthetap{i}{r_i}}=
\int_{0}^{+\infty} \cdots \int_{0}^{+\infty}\prod_{i=1}^n \xthetap{i}{r_i}f_{n}(\xb) \d{\xb},
\eeq
where, $f_{n}(\xb)$ is the multivariate Pareto (IV) density function (\ref{eq3}).
 This derivation is done in the following steps:\\
 First consider the following one dimensional integral:
\beqnx
C_{1}
&=&\int_{0}^{+\infty}\frac{\alpha}{\theta_1\gamma_1}\xthetap{1}{r_1}
{\pth{\frac{x_{1}}{\theta_{1}}}^
{\frac{1}{\gamma_{1}}-1}}\ppj^{-(\alpha+n)} \d{x}_{1} 
\\
&&
=\int_{0}^{+\infty}\frac{\alpha}{\theta_1\gamma_1}\xthetap{1}{r_1}
{\pth{\frac{x_{1}}{\theta_{1}}}^
{\frac{1}{\gamma_{1}}-1}\pth{1+\sum_{j=2}^n\xthetap{j}{1}}^{-(\alpha+n)}} 
\\
&&
\pth{1+\frac{\xthetap{1}{1}}{1+\sum_{j=2}^n\xthetap{j}{1}}}^{-(\alpha+n)} \d{x}_{1}.
\eeqnx
Note that, goings from first line to second line is just a
factorizing and rewriting the last term of the integral.
After many reflections on the links between Pareto (IV) and Burr families and
 Gamma and Beta functions, we found that the following
 change of variable
\beq
\pth{{1+\frac{\xthetap{1}{1}}{1+\sum_{j=2}^n\xthetap{j}{1}}}}=\frac{1}{1-t},\quad 0<t<1,
\eeq
simplifies this integral and guides us to the following result
\beq
C_{1}=\frac{\alpha\Gamma(r_{1}+1)\Gamma(\alpha+n-r_{1}-1)}{\Gamma(\alpha+n)}
\pth{1+\sum_{j=2}^n\xthetap{j}{1}}^{-(\alpha+n)+r_{1}+1}.
\eeq
Then we consider the following similar expression:
\beqnx
C_{2}
&=&\int_{0}^{+\infty}\frac{\alpha(\alpha+1)}{\theta_2\gamma_2}
\frac{\Gamma(r_{1}+1)\Gamma(\alpha+n-r_{1}-1)}{\Gamma(\alpha+n)}\xthetap{2}{r_2}
\\
&&
{\pth{\frac{x_{2}}{\theta_{2}}}^
{\frac{1}{\gamma_{2}}-1}}
\pth{1+\sum_{j=2}^n\xthetap{j}{1}}^{-(\alpha+n)+r_{1}+1}\d{x}_{2}  
\\
&&
=\int_{0}^{+\infty}
\frac{\alpha(\alpha+1)}{\theta_2\gamma_2}
\frac{\Gamma(r_{1}+1)\Gamma(\alpha+n-r_{1}-1)}{\Gamma(\alpha+n)}\xthetap{2}{r_2}
{\pth{\frac{x_{2}}{\theta_{2}}}^
{\frac{1}{\gamma_{2}}-1}}  
\\
&&
\pth{1+\sum_{j=3}^n\xthetap{j}{1}}^{-(\alpha+n)+r_{1}+1}
\pth{{1+\frac{\xthetap{2}{1}}{1+\sum_{j=3}^n\xthetap{j}{1}}}}^{-(\alpha+n)+r_{1}+1}
\d{x}_{2}, 
\eeqnx
and again using the following change of variable:
\beq
\pth{{1+\frac{\xthetap{2}{1}}{1+\sum_{j=3}^n\xthetap{j}{1}}}}=\frac{1}{1-t},
\eeq
we obtain:
\beqn
C_{2}
&=&\frac{\alpha(\alpha+1)\Gamma(r_{1}+1)\Gamma(r_{2}+1)\Gamma(\alpha+n-r_{1}-r_{2}-2)}{\Gamma(\alpha+n)}
\nonumber \\
&&
\pth{1+\sum_{j=3}^n\xthetap{j}{1}}^{-(\alpha+n)+r_{1}+r_{2}+2}.
\eeqn

Continuing this method, finally, we obtain the general expression:

\beq
C_{n}=\esp{\prod_{i=1}^n \Xthetap{i}{r_i}}=
 \frac{\Gamma(\alpha-\sum_{i=1}^n r_{i}) \prod_{i=1}^n
  \Gamma(r_{i}+1)}{\Gamma(\alpha)}, 
  \sum_{i=1}^n r_{i}< \alpha,\quad r_{i}>-1.  
\eeq
We may note that to simplify the lecture of the paper we did not give all
the details of these calculations.

\newpage
\bibliographystyle{amsplain}
\bibliography{mybib}

\end{document}